\documentclass[twocolumn,prl,showpacs,preprintnumbers,amssymb]{revtex4-1}

\usepackage{epsfig}
\usepackage{dcolumn}
\usepackage{bm}

\usepackage{amsmath}
\usepackage{amssymb}
\usepackage{latexsym}
\usepackage{epsfig}
\usepackage{color}
\usepackage{mathtools}
\usepackage{youngtab}
\usepackage{bigints}
\usepackage{sidecap}

\def\IC{\bf C}
\def\IZ{\bf Z}

\def\z2z2{$\IC^3/(\IZ_2\times\IZ_2)$}

\def\id{{\bf 1}}

\def\cG{\cal G}

\def\cp{\mbox{\bbbold C}\mbox{\bbbold P}}

\def\a{\alpha}

\def\b{\beta}

\def\d{\delta}\def\D{\Delta}

\def\k{\kappa}
\def\l{\lambda}

\def\p{\pi}

\def\s{\sigma}

\def\th{\theta}

\def\beq{\begin{equation}}\def\eeq{\end{equation}}
\def\beqa{\begin{eqnarray}}\def\eeqa{\end{eqnarray}}
\def\barr{\begin{array}}\def\earr{\end{array}}

\def\wt{\widetilde}

\def\ds {{\del \hspace{-6.4pt} \slash}\;}

 \let\br=\bigr

\def\bd{\begin{document}}
\def\ed{\end{document}}
\def\ba{\begin{array}}
\def\ea{\end{array}}
\def\bea{\begin{eqnarray}}
\def\eea{\end{eqnarray}}
\def\ft#1#2{{\textstyle{{\scriptstyle #1}\over {\scriptstyle #2}}}}
\def\fft#1#2{{#1 \over #2}}
\newcommand{\be}{\begin{equation}}
\newcommand{\ee}{\end{equation}}
\newcommand{\eq}[1]{(\ref{#1})}
\def\eqs#1#2{(\ref{#1}-\ref{#2})}
\def\det{{\rm det\,}}
\def\tr{{\rm tr}}
\newcommand{\ho}[1]{$\, ^{#1}$}
\newcommand{\hoch}[1]{$\, ^{#1}$}
\def\ra{\rightarrow}

\def\Xh{\hat{X}}
\def\ah{\hat{a}}
\def\xh{\hat{x}}
\def\yh{\hat{y}}
\def\ph{\hat{p}}
\def\G{{\cal G}}
\def\Dth{{\Delta_\th}}

\def\bk{{\bf k}}
\def\bx{{\bf x}}
\def\br{{\bf r}}
\def\tr{{\rm tr \,}}
\def\Tr{{\rm Tr \,}}
\def\diag{{\rm diag \,}}
\def\tg{{\rm tg \,}}
\def\NPB#1#2#3{Nucl. Phys. B {\bf #1} (19#2) #3}
\def\PLB#1#2#3{Phys. Lett. B {\bf #1} (19#2) #3}
\def\PLBold#1#2#3{Phys. Lett. {#1B} (19#2) #3}
\def\PRD#1#2#3{Phys. Rev. D {\bf #1} (19#2) #3}
\def\PRL#1#2#3{Phys. Rev. Lett. {\bf #1} (19#2) #3}
\def\PRT#1#2#3{Phys. Rep. {\bf #1} C (19#2) #3}
\def\MODA#1#2#3{Mod. Phys. Lett.  {\bf #1} (19#2) #3}
\def\ov{\overline}

\def\preal{{\rm Re\,}}
\def\pim{{\rm Im\,}}

\def\ds{\displaystyle}
\def\yzero{\smash{\hbox{$y\kern-4pt\raise1pt\hbox{${}^\circ$}$}}}
\def\p{\partial}
\def\a{\alpha}
\def\b{\beta}
\def\g{\gamma}
\def\d{\delta}
\def\beq{\begin{equation}}
\def\eeq{\end{equation}}
\def\beqa{\begin{eqnarray}}
\def\eeqa{\end{eqnarray}}
\def\Om{\Omega}
\def\om{\omega}
\def\th{\theta}
\def\vt{\vartheta}
\def\vphi{\varphi}
\def\-{\hphantom{-}}
\def\ov{\overline}
\def\s2{\frac{1}{\sqrt2}}
\def\wh{\widehat}
\def\wt{\widetilde}
\def\oh{\frac{1}{2}}
\def\tr{{\rm tr \,}}
\def\Tr{{\rm Tr \,}}
\def\diag{{\rm diag \,}}
\def\vac{|0 \rangle}
\def\vm{\relax{n_{\text{v}}}}
\def\cc{{\cal C}}
\def\ck{{\cal K}}
\def\ci{{\cal I}}
\def\cu{{\cal U}}
\def\cG{{\cal G}}
\def\cn{{\cal N}}
\def\cam{{\cal M}}
\def\cp{{\cal P}}
\def\ct{{\cal T}}
\def\cv{{\cal V}}
\def\cz{{\cal Z}}
\def\ch{{\cal H}}
\def\cf{{\cal F}}
\def\tv{\tilde v}
\def\Dsl{\,\raise.15ex\hbox{/}\mkern-13.5mu D} 
\def\IZ{Z\kern-.4em  Z}
\def\id{{\rm 1}}

\def\ti{\times}
\def\til{\tilde}
\def\eps{\epsilon}
\def\k{\kappa}
\def\A{\Arrowvert}
\def\cw{{\cal W}}
\def\G{\Gamma}
\def\car{{\cal R}}
\def\l{\lambda}
\def\raw{\rightarrow}
\def\Raw{\Rightarrow}
\def\inte{{\bf Z}}
\def\cpx{{\bf C}}
\def\real{{\bf R}}
\def\Lam{\Lambda}
\def\D{\Delta}
\def\cb{{\cal B}}
\def\ca{{\cal A}}

\begin{document}

\preprint{MAD-TH-15-01~ IFT-UAM/CSIC-15-007
}

\title{Widening the Axion Window via Kinetic and St\"uckelberg Mixings}

\author{Gary Shiu$^{1,2,3}$, Wieland Staessens$^{4}$, and Fang Ye$^{1,2}$}
\affiliation{\small\slshape  $^{1}$ Department of Physics, 1150 University Avenue, University of Wisconsin, Madison, WI 53706, USA \\
$^{2}$ Center for Fundamental Physics and Institute for Advanced Study, Hong Kong University of Science and Technology, Hong Kong \\
$^{3}$ University of Chinese Academy of Sciences, Beijing, China \\
     $^{4}$ Instituto de F\'isica Te\'orica UAM-CSIC, Cantoblanco, 28049 Madrid, Spain}
\begin{abstract}
We point out that kinetic and St\"uckelberg mixings that are generically present in the low energy effective action of axions can  significantly widen the window of axion decay constants. We show that an effective super-Planckian decay constant can be obtained even when the axion kinetic matrix has only sub-Planckian entries. Our minimal model involves only two axions, a St\"uckelberg $U(1)$ and a modest rank instanton generating non-Abelian group. Below the mass of the St\"uckelberg $U(1)$, there is only a single axion with a non-perturbatively generated potential. In contrast to previous approaches, the enhancement of the axion decay constant is not tied to the number of degrees of freedom introduced. We also discuss how kinetic mixings can lower the decay constant to the desired axion dark matter window.
 String theory embeddings of this scenario and their phenomenological features are briefly discussed.
\end{abstract}
\maketitle

\section{Introduction}

Axions or more generally axion-like particles are among the most recurrent extensions of the Standard Model. 
Their defining shift symmetry, originally proposed to solve the strong CP problem~\cite{Peccei-Quinn}, turns out to have far-reaching consequences in many other contexts in particle physics and cosmology. The axionic shift symmetry constrains how they couple to each other and to other matter perturbatively, namely, solely via derivative couplings. 
These properties of axions also make them an interesting candidate for dark matter and/or the inflaton.
Generic arguments in quantum gravity suggest~\cite{Abbott:1989jw}~that a continuous global
symmetry is at best perturbatively exact. 
Indeed, the continuous shift symmetry is broken to a discrete one through the coupling of axions to non-perturbative instantons, which in turn induce a potential (in particular, masses) for the axions. 
Much of the axion physics is dictated by the axion decay constant, which defines the periodicity of the canonically normalized axions.
Axionic couplings scale inversely with the axion decay constant, and their masses are determined by the axion decay constant and the non-perturbative scales involved. 
For example, the QCD axion can make up the cold dark matter of the universe if its decay constant lies within the window $ 10^{9}$ GeV $\leq f_{\rm QCD} \leq10^{12}$ GeV~\cite{Turner:1989vc}, while the non-perturbative potential for an axion can realize large field inflation~\cite{Freese:1990rb} if the associated decay constant exceeds the (reduced) Planck mass, i.e., $f_{\rm inf} > M_{Pl}$.  

Axions are ubiquitous in string theory, as they arise from dimensional reduction of higher form fields which appear generically in string compactifications.
Their shift symmetries originate from gauge symmetries in extra dimensions.
Although the origins of various string axions and their shift symmetries differ, careful studies surveying all known formulations of string theory~\cite{Svrcek:2006yi,Banks:2003sx} pointed to a universal upper bound for their axion decay constant $f$:
\begin{equation}
f \lessapprox \frac{g^2}{8 \pi^2} M_{Pl}, 
\label{bound}
\end{equation}
with $g$ the
coupling constant of the 
4D
non-Abelian gauge group to which the 
axion couples anomalously. 
This leads to the folklore that the string axion decay constant cannot exceed
the Planck scale~\cite{Banks:2003sx} and  at the same time reveals the tension to attain the QCD axion window~\cite{Choi:1985je,Svrcek:2006yi}.

In this paper, we propose a new mechanism to widen the range of axion decay constants in theories where the intrinsic axion field range is limited.
An implicit assumption behind the aforementioned upper bound is the absence of mixings among axions, i.e., the eigenbasis for the axion kinetic terms matches that of the instanton potential terms.
However, it is not uncommon for axions to mix kinetically and in the presence of St\"uckelberg $U(1)$ gauge fields, there are even further mixing effects. Thus, it is conceivable for the light axion that survives in the low energy theory to have a field range that differs significantly from what the original Lagrangian might suggest.

To explore the theoretically allowed window of axion decay constants, we considered the general multi-axion Lagrangian and found that the bound in eq.~(\ref{bound}) can be significantly relaxed when mixing effects are taken into account. We hasten to stress that although our motivation is partly string theoretical, our results apply generally to quantum field theories with multiple axions.
 More explicitly, for a system of $N$ axions $a^i$ charged under $M$ $U(1)$ gauge symmetries through St\"uckelberg couplings and coupling anomalously to $P$ non-Abelian gauge groups, the low energy effective action reads as follows, 
 \begin{widetext}
\begin{eqnarray}
{\cal S}^{\rm eff} &= &-  \bigintsss \left[ \sum _{\alpha,\,\beta =1}^{M} f_{\alpha \beta}F^{\alpha} \wedge \star_4 F^{\beta}  + \sum _{A=1}^P\frac{1}{g_{A}^{2}} \Tr(G^A\wedge \star_4 G^A) 
+ \frac{1}{2}\, \sum_{i,j=1}^N \cG _{ij} (d a^{i}-\sum _{\alpha =1}^{M}k^{i}_\alpha A^\alpha) \wedge \star_4 (d a^{j}-\sum _{\beta =1}^{M}k^{j}_\beta A^\beta) \right. \notag \\
&&\qquad \qquad \left.  -\frac{1}{8\pi ^2}\sum _{A=1}^P\left( \sum_{i=1}^N  r_{iA} a^i \right) \text{Tr} ( G ^A\wedge G^A )  -\frac{1}{8\pi ^2}\sum _{\alpha, \beta=1}^M\left( \sum_{i=1}^N  s_{i\alpha \beta} a^i \right) F ^\alpha \wedge F^\beta + \ldots \right] \label{Eq:CompleteGeneralAxionGauge}.
\end{eqnarray}
\end{widetext}

We choose the convention that the axions $a^i$ have a periodicity of $2 \pi$ 
and thus their decay constants are determined by the kinetic terms, ${\cal G}_{ij}$.
The matrix $f_{\alpha \beta}$ encodes the coupling constants of and possible mixing among the $U(1)$ gauge symmetries with gauge potential $A^\alpha$ and field strength $F^\alpha$. 
$G^A$ denotes the field strength of the strongly coupled non-Abelian gauge groups that generate instanton potentials.
The axion kinetic terms exhibit two types of mixing effects: mixing due to a non-diagonal metric ${\cal G}_{ij}$ on the axion moduli space and mixing due to St\"uckelberg couplings for charges $k^i_\alpha\neq 0$. 
An additional form of mixing arises 
as the axionic directions coupling anomalously to the non-Abelian gauge groups do not necessarily 
correspond to the eigenbasis for the potentials. This is expressed through the integer coefficients $r_{iA}$ and $s_{i\alpha \beta}$, for which at least two different coefficients are simultaneously non-vanishing. The anomalous couplings of the axions to the $U(1)$ gauge groups are included for completeness, but are not expected to contribute to the axion potentials due to the absence of $U(1)$ instantons in four dimensions~\cite{U1Dbrane}. The $\ldots$ denote the possible presence of chiral fermions 
and/or generalized Chern-Simons terms, required to ensure vanishing gauge anomalies~\cite{Aldazabal:2002py}.
Moreover, due to the presence of the chiral fermions and/or generalized Chern-Simons terms, the anomalous couplings of the axions to the gauge instantons remain $U(1)$ gauge invariant when the axions carry St\"uckelberg charges as shown explicitly in section~2.2.1~of~\cite{SSY}.


\section{Kinetic and St\"uckelberg mixings}
To highlight the mixing effects
among axions,
 it suffices to consider a minimal set-up with two axions ($N=2$), one Abelian 
 and one non-Abelian gauge groups ($M=P=1$).
 We can drop the indices $\alpha$ and $A$, and neglect the anomalous coupling of the axions to the $U(1)$ field strength. In order to identify the axionic direction eaten by the $U(1)$ gauge boson through the St\"uckelberg mechanism and to determine the correct axion decay constants, we have to perform a set of transformations ($SO(2)$ rotations and rescalings) diagonalizing the kinetic terms for the two-axion system. 
 A linear combination $\zeta$ of the axions $a^1$ and $a^2$ will form the longitudinal component of the massive $U(1)$ gauge boson with a St\"uckelberg mass:
\begin{equation} \label{Eq:MetricMixingMetricEigenvalues}
M_{st} = \sqrt{\lambda_- (k^-)^2 + \lambda_+ (k^+)^2} ,
\end{equation}
while the orthogonal linear combination $\xi$ remains uncharged under
this $U(1)$.
Here, $\lambda_\pm$ correspond to the eigenvalues of the axion moduli space metric ${\cal G}_{ij}$:
\begin{equation}
\lambda_\pm =  \frac{1}{2}\left[ ({\cal G}_{11} + {\cal G}_{22}) \pm \sqrt{4 {\cal G}_{12}^2 + ({\cal G}_{11}-{\cal G}_{22})^2} \right],
\end{equation}
and the charges $(k^+,k^-)$ correspond to the $U(1)$ charges in the axion eigenbasis diagonalising the metric ${\cal G}_{ij}$:
\begin{equation}
 k^+ =  \cos \frac{\theta}{2} k^1 + \sin \frac{\theta}{2} k^2, \quad k^- = \sin \frac{\theta}{2} k^1 - \cos \frac{\theta}{2} k^2.
\end{equation}
The continuous parameter $\theta \in \left[0, 2\pi \right]$ encodes the amount of axion mixing associated to a non-diagonal metric ${\cal G}_{ij}$ through the parametrisation:
\begin{equation}
\cos \theta = \frac{{\cal G}_{11} - {\cal G}_{22}}{\lambda_+ - \lambda_- }, \quad \sin \theta = \frac{ 2 {\cal G}_{12}}{\lambda_+ - \lambda_-}.
\end{equation}
For a diagonal axion metric ${\cal G}_{ij}$, i.e.~${\cal G}_{12}=0$, the $U(1)$ charges $(k^+,k^-)$ reduce to the original charges $(k^1,-k^2)$ as they appear in eq.~(\ref{Eq:CompleteGeneralAxionGauge}).

In the unitary gauge, the axion $\zeta$ 
is part of
 the massive $U(1)$ gauge boson, 
 and thus 
 only the anomalous coupling between the axion $\xi$ and the non-Abelian gauge group prevails~\cite{SSY}, yielding an effective axion decay constant:
{\small
\begin{equation}\label{Eq:AxionDecayConstant}
f_\xi = \frac{\sqrt{\lambda_+ \lambda_-} M_{st}}{ \cos \frac{\theta}{2} \left(  \lambda_+  k^+  r_2  + \lambda_- k^- r_1 \right) + \sin \frac{\theta}{2}  \left( \lambda_- k^- r_2 - \lambda_+  k^+  r_1  \right)}.
\end{equation}}
The axion decay constant $f_{\xi}$ exists purely due to the presence of non-vanishing St\"uckelberg couplings ($k^i\neq0$), irrespective of the occurrence of a non-diagonal metric ${\cal G}_{ij}$ on the axion moduli space. 
Upon integrating out the massive $U(1)$ gauge boson and the non-Abelian degrees of freedom, the gauge instanton background generates an axion-potential for $\xi$ of the usual cosine-type:
\begin{equation}\label{Eq:EffPotAxion}
V_{\rm eff}(\xi) = \Lambda^4 \left[ 1 - \cos \left( \frac{\xi}{f_{\xi}} \right) \right], 
\end{equation}
where $\Lambda$ is related to the characteristic energy scale of the condensate.  The process of integrating out the massive gauge boson occurs in two steps, as explained in detail in section~2.2.1~of~\cite{SSY}: first, the axion $\zeta$ is eliminated from the effective action by going to the unitary gauge for the gauge boson. In the unitary gauge, the anomalous coupling of the axion $\zeta$ to the non-Abelian gauge group vanishes as a consequence of the vanishing anomaly condition. The orthogonal axionic direction $\xi$ then forms the remaining axion that couples anomalously to the non-Abelian gauge instantons, explaining why the effective potential in (\ref{Eq:EffPotAxion}) solely depends on $\xi$. In a second step, the massive gauge boson is then integrated out by virtue of its equation of motion, giving rise to ($M_{st}$-suppressed) four-point interactions among the chiral fermions. In the end, the uncharged axion $\xi$ is the only field serving as the inflaton, ensuring that the inflationary motion occurs along a gauge-invariant path.


To explore the physical field range of the axion $\xi$, it suffices to determine how its axion decay constant scales with the continuous parameters in a specific region of the moduli space. For illustrative purposes, let us consider three regions of the parameter space where the axion decay constant takes super-Planckian values: 

\noindent {\bf Region 1:} for small kinetic mixing in the metric, i.e.~$\theta~\approx~0$, the decay constant~(\ref{Eq:AxionDecayConstant}) takes the form:
\begin{equation}\label{Eq:ADCRegion1}
f_{\xi} = \frac{\sqrt{{\cal G}_{11} {\cal G}_{22}} M_{st}}{k^1 r_2 {\cal G}_{11} - k^2 r_1 {\cal G}_{22}} = \frac{\sqrt{{\cal G}_{22}} \sqrt{(k^1)^2 + \varepsilon^2 (k^2)^2 }}{k^1 r_2 - \varepsilon^2 k^2 r_1},
\end{equation}
where the continuous parameter $\varepsilon^2\equiv {\cal G}_{22}/{\cal G}_{11}$ indicates the amount of isotropy between the metric eigenvalues. 
The decay constant takes trans-Planckian values in the region of the moduli space where the continuous parameter $\varepsilon$ asymptotes to
\begin{equation}
\varepsilon^2 \rightarrow \frac{k^1 r_2}{k^2 r_1}.
\end{equation}

\noindent {\bf Region 2:} for perfect isotropy between the diagonal entries of the metric, i.e.~$\varepsilon^2 = 1$, and a non-negligible  amount of kinetic mixing, i.e.~$\theta \approx \frac{\pi}{2}$, the $U(1)$ charges $(k^+,k^-)$ reduce to $(\frac{k^1+k^2}{\sqrt{2}},\frac{k^1-k^2}{\sqrt{2}})$. If we assume $r_1 = r_2$ for simplicity,
the decay constant can be simplified to:
\begin{equation}\label{Eq:ADCRegion2}
f_{\xi}
= \frac{ \sqrt{{\cal G}_{11}} \sqrt{1+\varrho^2}  \sqrt{ (k^1)^2 + (k^2)^2 + 2 k^1 k^2 \varrho^2 } }{|(k^1-k^2) r_1|  \sqrt{1-\varrho^2}},
\end{equation}
where the continuous parameter $\varrho^2 \equiv {\cal G}_{12}/{\cal G}_{11}$ measures the amount of kinetic mixing. In this moduli space region the decay constant reaches trans-Planckian values whenever the non-diagonal entries in the metric are of the same order as the diagonal ones, namely for: 
\begin{equation}\label{Eq:ADCLimit2}
\varrho^2 \rightarrow 1.
\end{equation}

\begin{figure}[h]
\centering
\begin{tabular}{c@{\hspace{0.4in}}c}
\includegraphics[width=3.2cm,height=3.2cm]{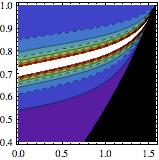} \begin{picture}(0,0) \put(-45,-10){$\theta$}  \put(-105,45){$\varepsilon$} \end{picture} 
& \includegraphics[width=3.2cm,height=3.2cm]{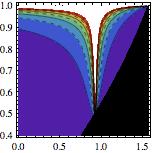} \begin{picture}(0,0) \put(-45,-10){$\theta$}  \put(-105,45){$\varepsilon$} \end{picture} 
\end{tabular}
\caption{Contour plots of decay constant $f_{\xi} (\theta, \varepsilon)$ for $2r_1=2r_2=2k^1=k^2 $ (left) and $r_1=2r_2=k^1=2k^2 $ (right). The $f_{\xi}$-values  range from small (purple) to large (red) following the rainbow contour colors. Unphysical regions with complex $f_{\xi}$ are located in the black band. \label{Fig:ContourPlots}}
\end{figure}
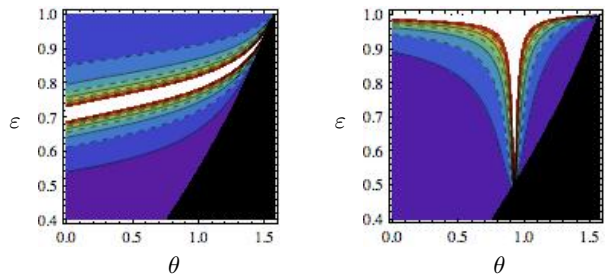

\noindent {\bf Region 3:} for intermediary kinetic mixing the range of the decay constant~(\ref{Eq:AxionDecayConstant}) can be represented through contour plots as functions of the continuous parameters $\varepsilon$ and $\theta \left(\in [0,\frac{\pi}{2}]\right)$ as in fig.~\ref{Fig:ContourPlots} upon fixing the $U(1)$ charges $k^i$ and the parameters $r_i$. Regions in the moduli space with $f_{\xi} > 10^2 \sqrt{{\cal G}_{11}}$ are highlighted in white. 

While we exploit multiple axions to obtain an effective super-Planckian decay constant, our mechanism differs fundamentally from earlier approaches.
Unlike N-flation \cite{Dimopoulos:2005ac} and aligned natural inflation \cite{Kim:2004rp}, 
the enhancement in the physical axion field range we found here is not tied to the number of degrees of freedom introduced (including axions, gauge fields, and any additional fields needed to ensure consistency of the theory). This can be seen already 
in the minimal setup above as an enhancement in 
neither (\ref{Eq:ADCRegion1}) nor (\ref{Eq:ADCRegion2}) 
requires adjusting the {\it discrete} parameters (e.g.~axion charges, axion-instanton couplings, and the rank of the non-Abelian gauge group) of the model but rather
{\it continuous} 
parameters (i.e. mixing angle $\theta$ and ratios $\varepsilon$ or $\varrho$ of metric entries) in the  axion moduli space
which leave the low energy spectrum intact.
This
decoupling of the axion field range enhancement 
from the low energy spectrum holds generally for the multi-axion system described by eq.~(\ref{Eq:CompleteGeneralAxionGauge}) and not just the minimal setup
considered here.
In contrast, 
 the enhancement in the axion field range 
 scales as 
 $\sim\sqrt{N}$ 
 in N-flation,
 and as $\sim \sqrt{N!}\,n^N$ \cite{Choi:2014rja}
 in aligned natural inflation, with
 $N$ the number of axions and
 $n \in \mathbb{Z}$ 
 the coefficients for the axion-instanton couplings. The presence of these light fields generically renormalize the Planck mass and we expect on general grounds~\cite{Dvali:2007hz} that $\delta M_{Pl}^2 \sim N$.
 Thus our scenario is minimal in
 that parametrically fewer degrees of freedom are needed to achieve the same enhancement and
so their associated quantum corrections to the Planck mass are less severe.

Let us end this section by briefly discussing the possibility
to lower the effective axion decay constant to
within the dark matter window.
If we consider the same configuration as in region 2, but assume that $r_1 = - r_2$ and $k^1 = k^2$, the axion decay constant instead reads:
\begin{equation}\label{Eq:ADCDarkMatter}
f_{\xi} = \frac{\sqrt{{\cal G}_{11}^2 - {\cal G}_{12}^2 }}{ \sqrt{2} |r_2 | \sqrt{ {\cal G}_{11} + {\cal G}_{12} }} = \frac{\sqrt{{\cal G}_{11}} \sqrt{1-\varrho^2}}{ \sqrt{2} |r_2 | },
\end{equation}
where the numerator decreases significantly in the limit~(\ref{Eq:ADCLimit2}). Considering moderate values for $r_2 \sim {\cal O}(1-10)$ and $\sqrt{{\cal G}_{11}} \sim {\cal O}(10^{15}-10^{17})$ GeV, a desired decay constant within the axion dark matter window can be obtained for moduli space regions with large kinetic mixing effects, i.e.~$1-\rho^2 \sim {\cal O}(10^{-4} - 10^{-8})$. More generically, eigenvalue repulsion can be used to lower the decay constant, similar to the $Z'$ masses considered in~\cite{Shiu:2013wxa,Feng:2014eja,Feng:2014cla}.  

In summary, by scanning the continuous moduli dependent parameter space for the axion moduli space metric ${\cal G}_{ij}$, we can find regions where the axion decay constant $f_\xi$ in eq.~(\ref{Eq:AxionDecayConstant}) takes trans-Planckian field ranges $f_\xi > M_{Pl}$ and regions where the decay constant falls within the classical axion decay window $10^9$ GeV $\leq f_\xi \leq 10^{12}$ GeV. These regions are mostly uncovered through the proposed kinetic mixing mechanisms in settings with a high amount of isotropy between the entries in the metric ${\cal G}_{ij}$. Nonetheless, the inclusion of kinetic mixing effects among axions allows for effective axion decay constants with a much broader energy window than the one of a single axion, alleviating the tension between current experimental bounds and the typical decay constants for string axions.

\section{String theory implementation}
It is natural to ask if our scenario can be realized in string theory where axion candidates are abundant. Axion models with a super-Planckian field range are sensitive to Planck scale physics. 
Thus, in such cases, a string theory implementation is not only natural but a necessity. Here we lay out the criteria that a string compactification needs to satisfy in order to implement the mechanisms we proposed above.
Closed string axions emerge naturally from the dimensional reduction of ten dimensional $p$-forms as summarised in table~\ref{tab:SummCSAII}, where 
for concreteness
we restricted to four-dimensional (4D)
Calabi-Yau (${\cal CY}_3$) orientifold compactifications of Type II superstring theory~\cite{Grimm:2004uq}. 
The background-dependence is reflected by the Hodge-numbers $h_\pm^{(1,1)}$, $h^{(2,1)}$ and $h^{(2,2)}_+$ expressing the number of orientifold-even/odd 2-forms, 3-forms and 4-forms respectively, and thereby setting the number $N_a$ of axions. De Rahm-duality then associates to every axion an orientifold-even/odd closed $p$-cycle $\gamma_i$ on the ${\cal CY}_3$ orientifold such that an axion $a^i$ can be defined as:
\begin{equation}
a^i \equiv \frac{1}{2\pi} \int_{\gamma_i} {\cal C}_{p}.
\end{equation} \begin{table}[h]
\begin{tabular}{|c||c|c||c|c|c|c|}
\hline \multicolumn{7}{|c|}{Overview of Type II closed string axions} \\
\hline
\hline &\multicolumn{2}{|c||}{Type IIA} & \multicolumn{4}{|c|}{Type IIB}\\
\hline
$p$-form ${\cal C}_p$&$B_2$ & $C_3$ & $C_0$ & $B_2$ & $ C_2$ & $C_4$\\
\hline
axion $a^i$ & $b^a$ & $\xi^k$& $c_0$ & $b^a$ & $c^a$ & $\rho^\alpha$   \\
$N_a$ & $h^{(1,1)}_-$ & $h^{2,1}+1$  & $1$ & $h_-^{(1,1)}$ & $h_-^{(1,1)}$ & $h_+^{(2,2)}$  \\
$U(1)$ &  & D6 on 3-cycle &  & & \multicolumn{2}{|c|}{D7 on 4-cycle} \\
\hline
\end{tabular}
\caption{Summary of model-dependent axions in Type II superstring theory on ${\cal CY}_3$ orientifolds~\cite{ORP:2015}. The D-brane configuration in the last row indicates which axions acquire St\"uckelberg $U(1)$ charges \label{tab:SummCSAII} and the origin of such $U(1)$'s.}
\end{table} 

\noindent The dimensional reduction of the kinetic terms for the $p$-forms yields the kinetic terms for the respective axions
whose continuous shift symmetries are remnants of the ten-dimensional gauge-invariance.
Furthermore, the kinetic terms for the axions are characterised by a non-diagonal metric ${\cal G}_{ij}$ on the axion moduli space, as in eq.~(\ref{Eq:CompleteGeneralAxionGauge}), except for the axion $c_0$. The metric ${\cal G}_{ij}$ depends on the moduli fields appearing in the same four-dimensional ${\cal N}=1$ supermultiplet as the respective axions. These moduli have to be stabilised at higher energy scales for the effective 
action
in eq.~(\ref{Eq:CompleteGeneralAxionGauge}) to be applicable.

The axions $\xi^k$, $c^a$ and $\rho^\alpha$ can be charged under the $U(1)$ gauge group~\cite{Grimm:2011dx,Jockers:2004yj} supported by the appropriate D$p$-brane as listed in table~\ref{tab:SummCSAII}. 
The St\"uckelberg couplings for the charged axions in 
eq.~(\ref{Eq:CompleteGeneralAxionGauge}) 
are
required for anomaly cancelation by virtue of the generalised Green-Schwarz mechanism though they can also appear for anomaly free $U(1)$'s. The axions $\xi^k$ and $c^a$ are charged under the $U(1)$ gauge symmetry when the corresponding D-brane wraps the $(6-p)$-cycle Poincar\'e dual  to the $p$-cycle associated with the axion. For the $\rho^\alpha$ axions to be charged under the Abelian gauge group, the D7-brane has to wrap the 4-cycle Poincar\'e dual to the 2-cycle supporting an
 internal magnetic 2-form flux. 
The
 St\"uckelberg charges $k^i$ 
 are thus directly related to the
integer wrapping numbers 
of the $U(1)$ D-brane
 along the internal dimensions.

The anomalous couplings to gauge instantons in eq.~(\ref{Eq:CompleteGeneralAxionGauge}) follow naturally from the reduction of the Chern-Simons action for the D-brane stack supporting the non-Abelian gauge group~\cite{Grimm:2011dx,Jockers:2004yj} . For the axions $\xi^k$ and $\rho^\alpha$ to couple anomalously to the non-Abelian gauge symmetry, it suffices that the corresponding D-brane stack wraps their associated $p$-cycles. For the $c^a$ axions the D-brane stack has to wrap the 4-cycle Poincar\'e dual to the 2-cycle supporting an internal magnetic 2-form flux.

Apart from gauge instantons, string theory also allows for the presence of D-brane instantons where Euclidean D-branes wrap $p$-cycles $\gamma_i$ on ${\cal CY}_3$ while being pointlike spacetime objects. The instanton amplitude is set by the action ${\cal S}_{E_{p-1}}$ for the Euclidean D-brane, which scales with the volume ${\rm Vol} (\gamma_i)$ of the wrapped $p$-cycle:
\begin{equation}
e^{- {\cal S}_{E_{p-1}} } = e^{-\frac{2\pi}{g_s} {\rm Vol}(\gamma_i) - i\, a^i }.
\end{equation}
The axion dependence in the phase then breaks the axion symmetry to the discrete shift symmetry $a^i \rightarrow a^i + 2 \pi$, implying that the moduli space for stringy axions is a torus $T^{N_a}$ equipped with metric ${\cal G}_{ij}$. Instanton corrections only contribute to the effective action when their fermionic zero modes can be saturated upon integration over the instanton moduli space, e.g.~for orientifold-invariant rigid cycles $\gamma_i$. For St\"uckelberg charged axions the D-brane instanton amplitude violates the $U(1)$ symmetry and effective contributions to the superpotential require the presence of $U(1)$ charged fermions whose collective charge cancels the $U(1)$ charge violation by the instanton to ensure gauge invariance~\cite{Blumenhagen:2006xt}. Such chiral fermions arise at the intersections of two D-branes in the bi-fundamental representation under the gauge groups supported by the respective D-branes. Which instanton type is the leading non-perturbative contribution and thereby sets the axion potential, is a model-dependent consideration. Explicit stringy realisations of set-up~(\ref{Eq:CompleteGeneralAxionGauge}) are constructed in~\cite{SSY} using Type IIA  with intersecting D6-branes \cite{Blumenhagen:2005mu} on the toroidal orientifold $T^6/\Omega{\cal R}$. A large axion decay constant is realised through certain isotropy relations among the complex structure moduli, analogous to the discussion for region~1~in~eq.~(\ref{Eq:ADCRegion1}).

\section{Conclusions}
In this paper, we propose and demonstrate
that
 kinetic and St\"uckelberg mixing effects
 can widen the axion window. 
 Our scenario applies generally to 
 field and string theories with multiple axions
 so long as the effective action in eq.~(\ref{Eq:CompleteGeneralAxionGauge}) is applicable.
In the context of string theory, our mechanism to lower the axion decay constant does not invoke large compact cycles, thereby
alleviating the requirement for an intermediate string mass scale~\cite{Cicoli:2012sz} or
the 
utility 
of field theory axions~\cite{Honecker:2013mya}. 
Our results thus open up new possibilities of detecting string axions through astrophysical, cosmological and laboratory means. It also allows to reconcile with a high fundamental string scale, should a detection of primordial tensor mode points us to high scale inflation.
On the other hand, an enhancement of the axion decay constant to super-Planckian values through mixings
enables us to realize ``natural-like inflation" in string theory. 
Generically, one expects 
the leading cosine potential (assumed to be exact in natural inflation~\cite{Freese:1990rb})  to receive model-dependent modifications from higher (or other subleading) instanton effects
when the effective axion decay constant becomes large~\cite{Comment}.
This expectation is in line with the weak gravity conjecture~\cite{ArkaniHamed:2006dz,Ooguri:2006in} whose formulation for
multi-axion systems is currently under investigation~\cite{BCSS}.
Nonetheless, the extended periodicity of the axion is not expected 
to be altered by these subleading corrections.
While such corrections 
  are model-dependent and hard to compute, their presence is suggestive of quantum gravity at work in order to couple the multiple axion system to gravity.
The deviation from a cosine potential 
could leave a measurable effect on the inflationary perturbation spectrum.
Quantifying 
such deviation 
requires a detailed understanding of the
ultraviolet completion of inflation
 and the moduli stabilization mechanism involved.
Turning this around, precise cosmological measurements
may point us closer to the structure of our string vacuum.

\subsection{Acknowledgments}

We thank Kiwoon Choi, Fernando Marchesano, Pablo Soler, and Angel Uranga for useful discussions.
The work of G.S. and F.Y. is supported in part by the DOE grant DE-FG-02-95ER40896 and the HKRGC grants HKUST4/CRF/13G, 604231 and 16304414. 
W.S.~is supported by the ERC Advanced Grant SPLE under contract ERC-2012-ADG-20120216-320421, by the grant FPA2012-32828 from the MINECO, and the grant SEV-2012-0249 of the ``Centro de Excelencia Severo Ochoa" Programme. W.S.~would also like to thank the European COST action MP1210 ``The String Theory Universe" for a Short Term Scientific Mission Grant.


\begin{thebibliography}{99}
\bibitem{U1Dbrane}
Note however that string theory allows for the existence of $U(1)$-type D-brane instantons. Such instantons can contribute to the effective superpotential when a Euclidean D-brane wraps the same internal cycle as the D-brane, even if the D-brane supports a $U(1)$ gauge group.
\bibitem{ORP:2015} 
  For Type IIB string theory, we considered the orientifold projection inverting the sign of the Calabi-Yau three-form. For an orientifold projection leaving the Calabi-Yau three-form completely invariant, the Hodge-numbers for the axions $c^a$ and $\rho^\alpha$ have to be swapped.
\end{thebibliography}
\end{document}